\documentclass[preprint]{jpsj2}

\renewcommand{\theenumi}{( \( \alph{enumi} \) )}
\def\runtitle{ Numerical Study for an Equilibrium
in the Recursive Stochastic State Selection Method }
\def\runauthor{Tomo  \textsc{Munehisa} and Yasuko \textsc{Munehisa}}

\title{%
 Numerical Study for an Equilibrium \\
in the Recursive Stochastic State Selection Method   }

\author{%
Tomo  \textsc{Munehisa} and
Yasuko \textsc{Munehisa}
}

\inst{%
 Faculty of Engineering, Univ. of Yamanashi, Kofu 400-8511
}
\recdate{\today}

\abst{%
We apply the recursive stochastic state selection method, 
which is a new method for Monte Carlo study we have recently developed,
to quantum spin systems with positive definite Hamiltonians.
Through numerical studies of two-dimensional $J_1-J_2$ Heisenberg model on 
a square lattice with unfrustrated couplings $J_1=1$ and $J_2=-1$ and
with non-frustrated ones $J_1=1$ and $J_2=0$, we find that 
a kind of equilibrium is realized in these systems. 
We also observe that in this equilibrium we can obtain a quite
accurate estimate of the energy eigenvalue for the system's ground
state. Statistical relative errors in our results are 
$0.03\%$ for the 36-site unfrustrated model and 
$0.06\%$ for the 64-site non-frustrated model. }

\kword{%

quantum spin, large size, numerical calculation, Monte Carlo,
square lattice, positive definite, equilibrium 
}
\begin{document}
\sloppy
\maketitle

\section{Introduction}

Quite recently we have proposed a new method for Monte Carlo
calculations to evaluate energy eigenvalues of quantum spin systems,
which we call the {\em stochastic state selection} (SSS) method\cite{mune}.
This method
enables us to obtain expectation values of powers of the Hamiltonian
even when limited computer memory resources are available. In the SSS
method we numerically select ``active'' basis states using {\em random choice
matrices}. A random choice matrix is a diagonal matrix whose elements are 
random variables to follow two-valued probability functions, 
named {\em on-off probability functions}, 
which are defined by an initial trial state of the system. 
Applying the method we succeeded in obtaining reliable values for energy
of the first excited states in the 64-site Shastry-Sutherland 
model\cite{mune2}.
Further, we have modified the method to develop the  
{\em recursive stochastic state selection} (RSSS) method, where 
the on-off probability functions depend on the intermediate states of the
system\cite{mune3}.
A merit of the RSSS method is
that errors of observed expectation values increase less rapidly 
than those in the SSS method when the power of the Hamiltonian rises.
We applied the RSSS method to the anti-ferromagnetic Heisenberg spin 
one-half system on the triangular lattice, which is a typical strongly 
frustrated system to which other methods such as the ordinary Monte
Carlo methods and the perturbative calculations are hardly applicable.
Our result for the lowest energy eigenvalue of the 36-site system is 
within one percent of the exact value\cite{mune3}.

In this paper we apply the RSSS method to quantum spin systems whose 
Hamiltonians are positive definite, namely whose Hamiltonian matrix 
elements are all non-negative. 
Tangible examples we employ here are 
the two-dimensional $J_1-J_2$ Heisenberg model on a square 
lattice\cite{szp,kosw} with unfrustrated or non-frustrated couplings.
Numerical studies suggest that in these
systems we are able to obtain, through repeated operations 
of the Hamiltonian and the random choice matrix,
a normalized state which contains a finite
and definite portion of the ground state. 
This implies a kind of equilibrium, which we call 
the {\em RSSS equilibrium}. Then we show that, by virtue of this equilibrium, 
one can easily estimate the energy eigenvalue of the ground state from the
normalization factor.

In the next section we briefly explain the RSSS method and give a  
definition of the RSSS equilibrium.
Section 3 adds an extended analytical discussion on models with
positive definite Hamiltonians, which indicate 
that they provide probable candidates to realize the RSSS equilibrium.
In sections 4 and 5 we investigate the $J_1-J_2$ model with $J_1=1$ and
$J_2 \leq 0$.  
In section 4 we present numerical results to examine the discussions in
sections 2 and 3. Here we demonstrate the realization of the RSSS equilibrium 
using small systems for which the exact ground states are calculable. 
First we study the non-frustrated model on a $4 \times 4 $ square
lattice setting $J_2 = -1$. Then the unfrustrated case ($J_2=0$) is 
investigated on the $6 \times 4 $ lattice.
Section 5 shows results on larger lattices, the $6 \times 6$ lattice with 
$J_2=-1$ and the $8 \times 8$ one with $J_2=0$. 
Here we describe how we obtain an estimate of the energy eigenvalue 
in the RSSS equilibrium. 
Our results thus obtained are 
$-40.644 \pm 0.013$ for the 36-site unfrustrated model, 
which is comparable to the value $-40.659$ obtained from ref.~4, 
and $-43.099 \pm 0.025$ for the 64-site non-frustrated model, 
which is only $0.02\%$ higher than $-43.107$\cite{square}, 
the most accurate value to our knowledge.   
The final section is devoted to summary and discussions.

\section{Recursive Stochastic State Selection Equilibrium}

First we briefly review the RSSS method\cite{mune3}.
Let us denote the Hamiltonian of a system by $\hat H$, a basis by $\{ \mid
i \rangle \}$, the number of the basis states by $N_{\rm V}$ and a trial
function by $\mid \psi^{(0)}\rangle $.  
We introduce the random choice matrices
$M_{\{ \eta ^{(m)}\}} = {\rm diag.} \{ \eta_1^{(m)}, \eta_2^{(m)},\cdots,  
\eta_{N_V}^{(m)}\}$ $(m=1,2,\cdots,L)$ in order to  calculate 
\begin{eqnarray}
  E_{\{\eta\}}(L) \equiv \langle \psi^{(0)} \mid \hat H
M_{\{ \eta ^{(L)}\}} \hat H M_{\{ \eta ^{(L-1)}\}} \cdots \hat H
M_{\{ \eta ^{(1)}\}} \mid  \psi^{(0)} \rangle ,
\label{eldef}
\end{eqnarray}
where the random variable $\eta_i^{(m)}$ is generated following 
the on-off probability function $ P_i^{(m)}(\eta)$.  In the RSSS method
we define
\begin{eqnarray}
 P_i^{(m)}(\eta) \equiv \frac{1}{a_i ^{(m)}} 
\delta \left( \eta - a_i ^{(m)} \right)
+ \left( 1 - \frac{1}{a_i ^{(m)}} \right) \delta \left( \eta \right) \ , 
\ \ \ \ \ \ \frac{1}{a_i ^{(m)}} 
\equiv min \left(1,\frac{|c_i^{(m-1)}|}{\epsilon} \right) \ ,
\label{pidefx}
\end{eqnarray}
using a positive parameter $\epsilon$ and 
the coefficient $c_i^{(m-1)}$ in the normalized intermediate state \\ 
$\mid \psi ^{(m-1)} \rangle = \sum \mid i \rangle c_i^{(m-1)}$
which is proportional to $\hat H M_{\{ \eta ^{(m-1)}\}} \cdots \hat H
M_{\{ \eta ^{(1)}\}} \mid  \psi^{(0)} \rangle$.
Starting from a given $\mid \psi^{(0)} \rangle$ we can recursively calculate  
\begin{eqnarray}
\mid \psi^{(m)} \rangle \equiv
\hat{H} M_{\{\eta ^{(m)} \}} \mid \psi^{(m-1)}\rangle / C^{(m)} \ ,
\label{psim} 
\end{eqnarray}
where $ C^{(m)} \ (>0)$ is the normalization factor calculated by
\begin{eqnarray}
\left[ C^{(m)} \right] ^2 &=& 
\langle \psi^{(m-1)} \mid  M_{\{\eta ^{(m)} \}}
 \hat{H}^2 M_{\{\eta ^{(m)} \}} \mid \psi^{(m-1)} \rangle \ .
\label{cmsquare}
\end{eqnarray}
We also define $\mid \chi ^{(m-1)} \rangle g^{(m-1)}$ by 
\begin{eqnarray}
\mid \chi^{(m-1)} \rangle \ g^{(m-1)} \equiv M_{\{ \eta ^{(m)} \}}
\mid \psi^{(m-1)} \rangle \ - \mid \psi^{(m-1)} \rangle = 
\sum_i \mid i \rangle c_i^{(m-1)} \left(\eta_i^{(m)} - 1 \right) 
\label{chimm1}
\end{eqnarray}
with the normalization condition 
$\langle \chi^{(m-1)} \mid \chi^{(m-1)} \rangle = 1$. It should be kept in
mind that for any state $\mid \Phi \rangle \equiv \sum \mid i \rangle b_i$ 
with $b_i$'s which are irrelevant to $\{\eta_i^{(m)}\}$, 
the statistical average of the inner product between $\mid \Phi \rangle$
and $\mid \chi^{(m-1)} \rangle g^{(m-1)}$ is zero,
\begin{eqnarray}
\langle \! \langle \ \langle \Phi \mid
\chi^{(m-1)} \rangle g^{(m-1)} \ \rangle \! \rangle 
= 0 \ ,
\label{sinn0}
\end{eqnarray}
because  
\begin{eqnarray}
\langle \! \langle \eta_i^{(m)} \rangle \! \rangle  
\equiv 
\int_0^{\infty} \eta_i^{(m)} P_i^{(m)}(\eta_i^{(m)}) d \eta_i^{(m)} = 1 \ .
\label{etaimsa}
\end{eqnarray}
Note that, as was discussed in ref.~3, 
this average should be calculated with a
fixed $\{ \{\eta^{(m-1)}\},\{\eta^{(m-2)}\},\cdots,\{\eta^{(1)}\} \}$.
In this paper we call such states as $\mid \chi^{(m-1)} \rangle
g^{(m-1)}$ {\em random states}.
For any random state $\mid \Omega \rangle$ which appears here we assume
that the deviation of $\langle \Phi \mid \Omega \rangle$ is negligibly
small so that     
$\langle \Phi \mid \Omega \rangle \ \simeq \ 0$
holds with no statistical average.  
%
Also remember that 
\begin{eqnarray}
\langle \! \langle \left[ g^{(m-1)}\right]^2 \rangle \! \rangle &=&
\langle \! \langle \sum_i \left[ c_i^{(m-1)}\right]^2
\left(\eta_i^{(m)}-1 \right)^2 \rangle \! \rangle = 
\sum_{0 < |c_i^{(m-1)}| < \epsilon } \left[ c_i^{(m-1)}\right]^2 
\left(\frac{\epsilon}{| c_i^{(m-1)}|} -1 \right)\nonumber \\
&=& \epsilon \sum_{|c_i^{(m-1)}| < \epsilon }|c_i^{(m-1)}| \ 
- \sum_{ |c_i^{(m-1)}| < \epsilon }\left[ c_i^{(m-1)}\right]^2 \ , 
\label{gmsquare}  
\end{eqnarray}
where we use (\ref{etaimsa}) and 
\begin{eqnarray}
\langle \! \langle \left[ \eta_i^{(m)} \right]^2 \rangle \! \rangle  
\equiv 
\int_0^{\infty} \left[\eta_i^{(m)}\right]^2
 P_i^{(m)}(\eta_i^{(m)}) d \eta_i^{(m)} = a_i^{(m)} \ .
\label{etaim2sa}
\end{eqnarray}

Now we discuss on the RSSS equilibrium. Let $\mid \psi_{\rm E} \rangle $
denote the exact eigenstate with the largest eigenvalue $E$.
We divide the intermediate state $\mid \psi^{(m)} \rangle $ into
a part which is proportional to $\mid \psi_{\rm E} \rangle $ and the
rest, 
\begin{eqnarray}
\mid \psi^{(m)} \rangle = \mid \psi_{\rm E} \rangle w^{(m)} + 
\mid \zeta ^{(m)} \rangle s^{(m)} \ , 
\label{sepes}
\end{eqnarray}
where 
\begin{eqnarray}
 w^{(m)} &\equiv& \langle \psi_{\rm E} \mid \psi^{(m)} \rangle \ , 
\label{defwm} \\
\mid \zeta ^{(m)} \rangle s^{(m)} &\equiv&  \mid \psi^{(m)} \rangle -
  \mid \psi_{\rm E} \rangle w^{(m)} \ , 
\label{defzmsm}
\end{eqnarray}
and we request $\langle \zeta ^{(m)} \mid \zeta ^{(m)} \rangle =1$.
Note that 
\begin{eqnarray}
\langle \psi_{\rm E}\mid \zeta ^{(m)} \rangle s^{(m)} = 0 \ , 
\label{pezmsm}
\end{eqnarray}
and 
\begin{eqnarray}
\left[w^{(m)}\right]^2 + \left[s^{(m)}\right]^2 = 1 \ ,  
\label{wm2sm2}
\end{eqnarray}
by definition.
What we mean by the RSSS equilibrium is that 
there exists a limit $w^{(eq)}$ defined by      
\begin{eqnarray}
\lim_{m_{\rm t} \rightarrow \infty} \frac{1}{m_{\rm t}} 
\sum_{m=m_{\rm s}}^{m_{\rm s}+m_{\rm t}-1}w^{(m)}= w^{(eq)} \ , 
\ \ \ \ \ (0 < w^{(eq)} \leq 1) \ , 
\label{wlim}
\end{eqnarray}
where $w^{(eq)}$ is independent of $m_{\rm s}$, whenever 
$m_{\rm s}$ is greater than or equal to some value of $m$, say $m_0$.

Let us derive a relation from this RSSS equilibrium for later use. 
Since from (\ref{psim}), (\ref{chimm1}) and (\ref{sepes})
\begin{eqnarray}
\mid \psi^{(m+1)} \rangle \ C^{(m+1)} &=& \hat H M_{\{\eta ^{(m+1)} \}} 
\mid \psi^{(m)} \rangle 
=\hat H \ \{ \ \mid \psi^{(m)} \rangle + 
\mid \chi^{(m)} \rangle g^{(m)} \  \}
\nonumber \\
&=& \hat H \ \{ \   \mid \psi_{\rm E} \rangle w^{(m)}
+ \mid \zeta^{(m)} \rangle \ s^{(m)} + \mid \chi^{(m)} \rangle g^{(m)} \  \}
\nonumber \\
&=& E  \mid \psi_{\rm E} \rangle w^{(m)} + \hat H  \mid 
\zeta^{(m)} \rangle \ s^{(m)} + \hat H \mid \chi^{(m)} \rangle g^{(m)} \ ,  
\label{psimp1}
\end{eqnarray}
we obtain, using (\ref{defwm}) and (\ref{pezmsm}),
\begin{eqnarray}
w^{(m+1)}  C^{(m+1)}= 
\langle \psi_{\rm E} \mid \psi^{(m+1)} \rangle\ C^{(m+1)}  =
Ew^{(m)}+ E \langle  \psi_{\rm E} \mid \chi^{(m)} \rangle g^{(m)} \ ,   
\label{wmp1ex}
\end{eqnarray}
which leads 
\begin{eqnarray}
w^{(m+1)} = \frac{E}{ C^{(m+1)}} \  w^{(m)} \ ,
\label{wmp1sim}
\end{eqnarray}
if the second term in the right-hand side of (\ref{wmp1ex}) is negligible.
Then we can expect 
\begin{eqnarray}
E = C^{(m+1)} \ ,  
\label{estime} 
\end{eqnarray}
for sufficiently large values of $m$. Thus we become aware that 
the value of $E$ can be estimated by the normalization factor 
$C^{(m+1)}$.
Actually, contributions from the second term in (\ref{wmp1ex}) 
should be taken into account when we consider the fluctuation of $C^{(m+1)}$.

\section{RSSS Equilibrium for Positive Definite Hamiltonians}

In this section we deal with several quantities  
concerning to the systems with positive definite Hamiltonians, 
which will be referred to in our numerical work. 
After the manner of the previous section we make our discussion as
simple as possible, neglecting all fluctuations in these quantities. 

In order to make our analysis concrete, we here limit
ourselves to a case where 
\begin{eqnarray}
\mid \psi^{(0)} \rangle = \mid \psi_{\rm E} \rangle
\equiv \sum_i \mid i \rangle f_i \ \ \ \ \ \ 
(f_i \geq 0 \ {\rm for \ all \ }i) \ . 
\label{inipsi}
\end{eqnarray}
We suppose we have chosen an adequate basis 
$\{\mid i \rangle \}$ so that all $f_i$'s as well as 
$h_{ij} \equiv \langle i \mid \hat H \mid j \rangle$'s are 
non-negative, which is always possible for positive definite Hamiltonians.
It should be noted that all coefficients in the expansion of 
$\mid \psi ^{(1)} \rangle$,  $\mid \psi ^{(2)} \rangle$, $\cdots$ are 
also non-negative in this case, 
because in the relation between $c_i^{(m)}$ and $c_j^{(m-1)}$,   
\begin{eqnarray}
c_i^{(m)} = \sum_j  h_{ij} c_j^{(m-1)} \eta_j^{(m)} / C^{(m)} \ ,
\label{cim}
\end{eqnarray}
which we learn from (\ref{psim}), $h_{ij} \geq 0$ for all $i$ and $j$ 
and $\eta_j^{(m)} /C^{(m)} \geq 0$ for all $j$ by definition.

We pay our attention to a relation led from (\ref{psimp1}) which defines
$\mid \psi^{(m+1)} \rangle C^{(m+1)}$, 
\begin{eqnarray}
\left[ C^{(m+1)}\right]^2 &=& E^2  \left[ w^{(m)}\right]^2 + 
\langle \zeta^{(m)}\mid \hat H ^2 \mid \zeta^{(m)} \rangle 
\left(1- \left[ w^{(m)}\right]^2 \right) 
+ \langle \chi^{(m)}\mid \hat H^2 \mid \chi^{(m)} \rangle 
\left[ g^{(m)}\right]^2 \nonumber \\  
&+& 2 E^2 w^{(m)} \langle \psi_{\rm E} \mid \chi^{(m)} \rangle g^{(m)}    
+ 2 s^{(m)} \langle \zeta^{(m)}  \mid \hat H^2 \mid 
\chi^{(m)} \rangle g^{(m)} 
\ ,
\label{cmp1sq}
\end{eqnarray}
where we used (\ref{pezmsm}) and (\ref{wm2sm2}).
For sufficiently large value of $m$ we expect that both
$\langle \zeta^{(m)} \mid \hat H^2 \mid \zeta^{(m)} \rangle$ and 
$\langle \chi^{(m)} \mid \hat H^2 \mid \chi^{(m)} \rangle$ are
independent of $m$, 
 \begin{eqnarray}
\langle \zeta^{(m)} \mid \hat H^2 \mid \zeta^{(m)} \rangle \ &\simeq& \
H_{2\zeta} \ , \label{as1} \\
\langle \chi^{(m)} \mid \hat H^2 \mid \chi^{(m)} \rangle \ &\simeq& \
H_{2\chi} \ , \label{as2} 
\end{eqnarray}
where $H_{2\zeta}$ and $H_{2\chi}$ denote positive constants determined 
by $\epsilon$ and $\hat H$. 
We also expect both cross terms in (\ref{cmp1sq}) are 
negligible because statistical averages of them vanish\cite{foot4}. 
Validity of these assumptions will be numerically examined in the next 
section. We thus obtain a relation
\begin{eqnarray}
\left[ C^{(m+1)}\right]^2 \ \simeq \ 
E^2 \left[w^{(m)}\right]^2 + H_{2\zeta}\left( 1- \left[w^{(m)}\right]^2\right)
+H_{2\chi} \left[ g^{(m)}\right]^2 \ . 
\label{wright}
\end{eqnarray}
Let us examine $\left[g^{(m)}\right]^2$ then. Using (\ref{gmsquare}) we obtain 
\begin{eqnarray}
\langle \! \langle \left[ g^{(m)}\right]^2 \rangle \! \rangle 
+ \sum_{|c_i^{(m)}| < \epsilon } \left[c_i^{(m)} \right]^2 \ = \ 
\epsilon \sum_{|c_i^{(m)}| < \epsilon }|c_i^{(m)}| \ \leq \ 
\epsilon \sum_i |c_i^{(m)}| = \epsilon \sum_i c_i^{(m)} \ . 
\label{gmsup}
\end{eqnarray}
The last equality follows from the fact that all $c_i^{(m)}$ are
non-negative here. Note that 
$ \sum_{|c_i^{(m)}| < \epsilon } \left[c_i^{(m)} \right]^2 \ \leq \ 1$ 
because of the normalization condition. 
For $m=1$ we obtain from (\ref{cim}) 
\begin{eqnarray}
c_i^{(1)} = \sum_j  h_{ij} f_j \eta_j^{(1)} / C^{(1)} 
= f_i E/ C^{(1)} + \sum_j  h_{ij} f_j \left(\eta_j^{(1)}-1\right)/ C^{(1)} \ ,
\label{ci1}
\end{eqnarray}    
using the relation $\sum  h_{ij} f_j = E f_i$ which
results from $\hat H \mid \psi_{\rm E} \rangle = E \mid \psi_{\rm E}
\rangle$. Therefore
\begin{eqnarray}
\epsilon \sum_i c_i^{(1)} \ \sim \
 w^{(1)} \epsilon \sum_i f_i + \epsilon \sum_i \sum_j  h_{ij} f_j 
\left(\eta_j^{(1)}-1\right)/ C^{(1)} \ ,
\label{g1supp}
\end{eqnarray}
where we use (\ref{wmp1sim}), notifying that $w^{(0)}=1$ for (\ref{inipsi}). 
We assume that the second term in the right-hand side of (\ref{g1supp}) 
is negligible because sums over $i$ and $j$ in it would be enough to promote 
the cancellation in $\left(\eta_j^{(1)}-1\right)$ keeping $ C^{(1)}$
almost irrelevant to the sampling. Therefore we acquire a relation 
$\epsilon \sum_i c_i^{(1)} \ \simeq \ w^{(1)} \epsilon \sum_i f_i$. 
In the same manner we obtain 
\begin{eqnarray}
c_i^{(2)} = \sum_j  h_{ij} c_j^{(1)} \eta_j^{(2)} / C^{(2)} 
\ \sim \ w^{(2)} f_i  + \sum_j \sum_l h_{ij}  h_{jl} f_l  
\left(\eta_j^{(2)}\eta_l^{(1)}-1\right)/ C^{(2)}C^{(1)} \ ,
\label{ci2}
\end{eqnarray}
which leads, taking  $\langle \! \langle \ \langle \! \langle
\left(\eta_j^{(2)}\eta_l^{(1)}-1\right) \rangle \! \rangle _{{\rm fixed}
\{\eta^{(1)}\}} \rangle \! \rangle = 0$ into account, 
$\epsilon \sum_i c_i^{(2)} \ \simeq \ w^{(2)} \epsilon \sum_i f_i  $.
Thus we expect  
\begin{eqnarray}
\langle \! \langle \left[ g^{(m)}\right]^2 \rangle \! \rangle
\ \lesssim \ w^{(m)} \epsilon \sum_i f_i 
- \sum_{|c_i^{(m)}| < \epsilon }\left[c_i^{(m)} \right]^2 \ , 
\label{gmsupp}
\end{eqnarray}
holds for any $m$.
In order to propose a relation to be expected in the RSSS equilibrium
we add, based on the above discussions, 
the following assumption for sufficiently large $m$, 
\begin{eqnarray}
\sum_{|c_i^{(m)}| < \epsilon } \left[c_i^{(m)} \right]^2 \ &\simeq& \ 
K \ , \label{as3} \\
\left[ g^{(m)} \right]^2 \ &\simeq& \ G w^{(m)} - K \ , \ \ \ \ \ G \equiv 
\epsilon \sum_i f_i  \ , 
\label{as4}
\end{eqnarray}
where $K$ $(0 < K \leq 1)$ is a constant defined by $\epsilon$. 
We will see that these assumptions are acceptable for systems we
numerically study.  

Now we reach to a function which determines $C^{(m+1)}$ from $w^{(m)}$,  
\begin{eqnarray}
\left[ C^{(m+1)}\right]^2 \ &\simeq& \ 
E^2 \left[w^{(m)}\right]^2 + H_{2\zeta} 
\left( 1- \left[w^{(m)}\right]^2 \right)+  
 H_{2\chi} \left( G w^{(m)} - K \right) \nonumber \\  
&=& \left(E^2- H_{2\zeta}\right)\left[w^{(m)}\right]^2 + G H_{2\chi} w^{(m)} + 
\left(H_{2\zeta} - K H_{2\chi} \right) \ .
\label{cmp1fnc}
\end{eqnarray}
Together with (\ref{wmp1sim}) this leads to a relation between
$w^{(m+1)}$ and $w^{(m)}$, 
\begin{eqnarray}
w^{(m+1)} \ \simeq \ 
 \frac{Ew^{(m)}}{\sqrt{\left(E^2- H_{2\zeta}\right)\left[w^{(m)}\right]^2 
+ G H_{2\chi} w^{(m)} + \left(H_{2\zeta} - K H_{2\chi} \right)}} \ .
\label{wmp1fnc}
\end{eqnarray}
Next notify it is mathematically guaranteed that when we define $\{x_n\}$ by
\begin{eqnarray}
x_{n+1}= f(x_n) \ ,\ \ \ 0 < x_0 \leq 1 \ ,
\label{xnp1fnc}
\end{eqnarray}
with  
\begin{eqnarray}
f(x) \equiv \frac{px}{\sqrt{ax^2+2bx+c}} \ \ \ (0 \leq x \leq 1) \ ,
\label{fx}
\end{eqnarray}
the value of $x_n$ converges to a {\em finite} value between 0 and 1 
provided $a > 0$, $b > 0$, $c > 0$, $p > 0$ and $p/\sqrt{a+2b+c} < 1$.
The reason for it is that, under these conditions, $f(0)=0$, $f(1) < 1$ and  
\begin{eqnarray}
f''(x)= -p \ \frac{2abx^2+(b^2+3ac)x+2bc}{\left[\sqrt{ax^2+2bx+c}\ \right]^5} 
< 0 \ .
\label{reasons}
\end{eqnarray}
Let $a=E^2- H_{2\zeta}$, $b=G H_{2\chi}/2$, $c=H_{2\zeta} - K H_{2\chi}$
and $p=E$ then in (\ref{wmp1fnc}). 
Since $E$ is the largest eigenvalue of the positive
definite $\hat H$, it is clear that $a > 0$, $b > 0$ and $p > 0$ 
by definition. 
In addition, we see that $a+2b+c =
\left(E^2- H_{2\zeta}\right) + G H_{2\chi} + \left(H_{2\zeta} - 
K H_{2\chi} \right) = E^2 + H_{2\zeta}\left( G-K \right) > E^2 = p^2 $ 
because we obtain 
$G-K \geq Gw^{(m)}-K \sim \left[ g^{(m)} \right] ^2 > 0 $ using (\ref{as4}). 
Therefore (\ref{wmp1fnc}) suggests that the RSSS equilibrium will be
realized when $H_{2\zeta} > K H_{2\chi}$, which seems mostly fulfilled 
irrespective of values of $\epsilon$\cite{foot5}. 

Finally let us point out that we can obtain an equation for $w^{(eq)}$ 
combining (\ref{estime}) and (\ref{cmp1fnc}), which is 
\begin{eqnarray}
\left(E^2- H_{2\zeta}\right) \left[w^{(eq)}\right]^2 + G H_{2\chi}  w^{(eq)} - 
\left(E^2 - H_{2\zeta} + K H_{2\chi} \right) = 0 \ .
\label{equcw}
\end{eqnarray}
The relevant solution is
\begin{eqnarray}
w^{(eq)} = -q + \sqrt{q^2 + 1+ \kappa } \ , \ \ \  
q \equiv \frac{1}{2} 
\cdot \frac{GH_{2\chi}}{E^2 -  H_{2\zeta} } \ ,
\ \ \ \kappa \equiv \frac{K H_{2\chi}}{E^2 -  H_{2\zeta} }
\ ,
\label{solcw}
\end{eqnarray}
which should be compared with $w^{(m)}$ 
to examine the validity of (\ref{cmp1fnc}).

\section{Numerical Study on Small Systems}
 
Here we numerically study two quantum spin one-half systems on
small lattices in order to examine assumptions and
relations discussed in previous sections. 

First we investigate the $4 \times 4$ $J_1-J_2$ model with couplings 
$J_1=1$ and $J_2=-1$.
This system size is too small to satisfactorily suppress fluctuations 
of the observed data for $\epsilon \gtrsim 0.05$, 
yet it has a merit that we can easily obtain the exact ground state 
by means of the Lanczos method.
The Hamiltonian of the system is  
\begin{eqnarray}
\hat H_{J_1 J_2} \equiv 
\frac{J_1}{4} \sum_{(nn)} \mbox{\boldmath $\sigma$}_i \cdot 
\mbox{\boldmath $\sigma$}_j + 
\frac{J_2}{4} \sum_{(nnn)} \mbox{\boldmath $\sigma$}_i \cdot
\mbox{\boldmath $\sigma$}_{j'}  \ ,
\label{j1j2h}
\end{eqnarray}
where the first summation is over the nearest neighbor pairs and the
second one over the next-nearest neighbor pairs on the square 
lattice\cite{kosw}.
In the following calculations we employ, instead of $\hat H_{J_1 J_2}$,
\begin{eqnarray}
\hat H \equiv l \hat I + U^{\dag} \left(- \hat H_{J_1 J_2} \right)U \ ,
\label{j1j2q4}
\end{eqnarray}
where $\hat I$ is the identity operator and $U$ denotes a unitary 
transform to make off-diagonal elements $h_{ij} \ (i \neq j)$
non-negative. The value of the parameter $l$ is determined so as to it 
ensures that all diagonal elements of $\hat H$ are non-negative. 
We put $l=8$ here.
Let $\mid \psi_{\rm E} \rangle$ be the eigenstate for
the largest eigenvalue of $\hat H$, which is expressed by $E$, as we did in
the previous section. 
The number of basis states for $\mid \psi_{\rm E} \rangle$ is $12,870$
and we obtain the exact value of $E$ for this system is $26.633$, 
which is in agreement with the result reported in ref~4. We denote the
exact value of $E$ by $E_0$ hereafter. Note that,  
with this positive definite $\hat H$, all coefficients in the 
expansion of $\mid \psi_{\rm E} \rangle$ are non-negative.   
Starting $\mid \psi_{\rm E} \rangle$ we calculate $\mid \psi^{(m)}
\rangle$ up to $m=1000$ using the $k$-th random number sequence 
($1 \leq k \leq n_{\rm smpl}$) to generate $ \{\eta\} \equiv
\{ \{\eta^{(m)}\}, \{\eta^{(m-1)}\}, \cdots, \{\eta^{(1)}\} \}$, 
which we refer to as {\it a $rns$-sample} hereafter.
Although the initial transient process terminates up to $m \sim 5$,
we need to carry out such long runs because some quantities measured with 
$\epsilon \gtrsim 0.05$ show much fluctuations ranging a few hundreds of $m$. 
Figures 1 - 7 show results from one $rns$-sample. 
The number of basis states whose coefficients are non-zero is saturated
to be $\sim 5 \times 10^2$ ($\sim 1.3 \times 10^3$, $\sim 3.8\times 10^3$) 
in the expansion of
$M_{\{\eta ^{(m)} \}} \mid \psi^{(m-1)}\rangle$ for $\epsilon = 0.1$
(0.05, 0.02), while it is 
$\sim 8.5\times 10^3$ ($\sim 1.2 \times 10^4$, $\sim 12,870$) 
in the expansion of
$\hat H M_{\{\eta ^{(m)} \}} \mid \psi^{(m-1)}\rangle$.  
The results are not much different for other $rns$-samples.

Figures 1 - 4 are to numerically examine the assumptions employed when 
we obtain (\ref{wright}) from (\ref{cmp1sq}).  
Figures 1, 2 and 3 plot quantities which appear in the right-hand side of 
(\ref{cmp1sq}) obtained for $\epsilon = 0.1$, $0.05$ and $0.02$,
respectively.
We see that cross terms 
$2 E^2 w^{(m)} \langle \psi_{\rm E} \mid \chi^{(m)} \rangle g^{(m)} $    
and $2 s^{(m)} \langle \zeta^{(m)}  \mid \hat H^2 \mid 
\chi^{(m)} \rangle g^{(m)} $ 
distribute around $0$ and are negligible compared to other three terms
$E^2\left[w^{(m)}\right]^2$, 
$\langle \zeta^{(m)}\mid \hat H ^2 \mid \zeta^{(m)} \rangle
\left(1-\left[w^{(m)}\right]^2\right)$ and
$\langle \chi^{(m)}\mid \hat H ^2 \mid \chi^{(m)} \rangle
\left[g^{(m)}\right]^2$.
Figure~4 shows $\langle \zeta^{(m)} \mid \hat H^2 \mid \zeta^{(m)}
\rangle $ and $\langle \chi^{(m)} \mid \hat H^2 \mid \chi^{(m)}\rangle $
calculated from $\mid \zeta^{(m)} \rangle $ defined in (\ref{defzmsm})
and $\mid \chi^{(m)} \rangle $ defined in (\ref{chimm1}).
Although there are much fluctuations in data for 
$\langle \zeta^{(m)} \mid \hat H^2 \mid \zeta^{(m)}\rangle $, our
assumption in (\ref{as1}) that they are nearly equals to a constant 
$H_{2\zeta}$ seems acceptable at least when $\epsilon = 0.02$.    
As for $\langle \chi^{(m)} \mid \hat H^2 \mid \chi^{(m)}\rangle $, we 
see the data are stable to support that (\ref{as2}), namely 
$\langle \chi^{(m)} \mid \hat H^2 \mid \chi^{(m)}\rangle \simeq
 H_{2\chi}$, holds. 
Note that the condition $H_{2\zeta} > K H_{2\chi}$ ($0 < K \leq 1$) 
is guaranteed from these data. 

In fig.~5 we plot $ \sum_{|c_i^{(m)}| < \epsilon } \left[c_i^{(m)}
\right]^2$ together with $K$ which is estimated from the average of 
them with $m=800, \cdots,999$.     
Figure~6 compares $\left[g^{(m)}\right]^2$ calculated from (\ref{chimm1}) 
with $G w^{(m)} -K$ $(G \equiv \epsilon \sum f_i)$, where we calculate 
$w^{(m)}$ from (\ref{defwm}) and use $K$ obtained in Fig.~5.
These results suggest that the assumptions (\ref{as3}) and (\ref{as4})
are probably good for $\epsilon = 0.05$ and $\epsilon = 0.02$, although
data for $\epsilon = 0.05$ fluctuates a little and 
$G w^{(m)} -K$ for $\epsilon = 0.02$ slightly overestimates 
$\left[g^{(m)}\right]^2$. Results for $\epsilon = 0.1$, on the contrary,
are not fully convincing because of their fluctuations.   
In Fig.~7 we present data for $w^{(m)}$ together with the solution 
$w^{(eq)}$ in (\ref{solcw}), where we substitute the average of 
$ \sum_{|c_i^{(m)}| < \epsilon } \left[c_i^{(m)} \right]^2$,
$\langle \zeta ^{(m)} \mid \hat H ^2 \mid \zeta ^{(m)} \rangle$ and 
$\langle \chi ^{(m)} \mid \hat H ^2 \mid \chi ^{(m)} \rangle$ over 
$m=800,\cdots,999$ for $K$, $H_{2\zeta}$ and $H_{2\chi}$, respectively.
We see the agreement is satisfying. 
In order to make a numerical examination of (\ref{wlim}) 
we calculate, changing $m_{\rm s}$, averages of $w^{(m)}$'s
for finite values of $m_{\rm t}$. 
The results prove to be almost independent of $m_{\rm s}$.
Although data with $\epsilon = 0.1$ give us   
a little fluctuating results such as
$0.575$ $(m_{\rm s}=200, \ m_{\rm t}=800)$ and 
$0.584$ $(m_{\rm s}=800, \ m_{\rm t}=200)$, others are very stable.  
They are, for instance,
$0.7885$ $(m_{\rm s}=200, \ m_{\rm t}=800)$,
$0.7883$ $(m_{\rm s}=600, \ m_{\rm t}=400)$,
$0.7885$ $(m_{\rm s}=600, \ m_{\rm t}=200)$ and
$0.7881$ $(m_{\rm s}=800, \ m_{\rm t}=200)$ with $\epsilon = 0.05$ and
$0.9415$ $(m_{\rm s}=200, \ m_{\rm t}=800)$,
$0.9416$ $(m_{\rm s}=600, \ m_{\rm t}=400)$,
$0.9417$ $(m_{\rm s}=600, \ m_{\rm t}=200)$ and
$0.9415$ $(m_{\rm s}=800, \ m_{\rm t}=200)$ with $\epsilon = 0.02$.
From these results we conclude that 
the RSSS equilibrium is established in this system for all values of
$\epsilon $, and that our discussion to lead (\ref{solcw}) is valid.

Figure~8 plots 
\begin{eqnarray}
\langle \! \langle C^{(m)} \rangle \! \rangle _{n_{\rm smpl}}
 \equiv \frac{1}{n_{\rm smpl}}
\sum_{k=1}^{n_{\rm smpl}} C^{(m)}_{\{\eta\}_k} \ ,
\label{cmsmpl}    
\end{eqnarray}
where $C^{(m)}_{ \{\eta\}_k}$ denotes the normalization factor
calculated from (\ref{cmsquare}) for the $k$-th $rns$-sample. 
The total number of these samples is $n_{\rm smpl} = 100$.
The error is estimated by 
\begin{eqnarray}
   Er^{(m)} &\equiv& 2 \sqrt \frac{
\langle \! \langle \left[ C^{(m)}\right]^2 \rangle \! \rangle _{n_{\rm smpl}}-
\left[ \langle \! \langle C^{(m)} \rangle \! \rangle_{n_{\rm smpl}} \right]^2 }
{n_{\rm smpl}-1} \ .
\label{errsmpl}
\end{eqnarray} 
The exact value of $E$ is also shown in the Figure by a dashed line.
We observe that data for $\epsilon = 0.02$ support the approximate relation 
\begin{eqnarray}
E = \langle \! \langle C^{(m+1)} \rangle \! \rangle \ ,
\label{enecmav}    
\end{eqnarray}
which follows from (\ref{estime}). 
Data with $\epsilon = 0.05$ and $\epsilon = 0.1$
fluctuate much more than those with $\epsilon = 0.02$, but we can still see 
they are close to the exact value of $E$.
These fluctuations of $\langle \! \langle C^{(m+1)} \rangle \! \rangle$ 
would be mainly ascribed to the fact that 
the second term in (\ref{wmp1ex}),  which we evaluate to be    
$|\langle \psi_{\rm E}\mid \chi^{(m)}\rangle g^{(m)}/w^{(m)}| < 0.03$
($0.07$, $0.14$) for $\epsilon = 0.02$ ($0.05$, $0.1$) in our measurement,
is not completely negligible. We think this term is responsible for 
the $\epsilon$-dependency of 
$\langle \! \langle C^{(m+1)} \rangle \! \rangle $
which will be discussed in the following section.

Now let us show results for the $J_2=0$ case studied on a
$6 \times 4$ square lattice. Here we employ $l=6$ in (\ref{j1j2q4}), for
which $E_0$, the exact value of $E$, becomes $22.553$.
Although it is more time-consuming to investigate
this lattice size, we can obtain results which fluctuate less  
compared to those for the 16-site $J_2=-1$ model. 
The number of the basis states in the the exact eigenstate is
$2,704,157$. 
The number of basis states whose coefficients are non-zero is 
$\sim 2.4 \times 10^3$ ($\sim 4.5 \times 10^4$) in the expansion of
$M_{\{\eta ^{(m)} \}} \mid \psi^{(m-1)}\rangle$ for $\epsilon = 0.05$
(0.01), while it is $\sim 6.5 \times 10^4$ ($\sim 8.6 \times 10^5$) 
in the expansion of
$\hat H M_{\{\eta ^{(m)} \}} \mid \psi^{(m-1)}\rangle$.  

Figure~9 presents data of $\left[g^{(m)}\right]^2$ with $\epsilon=0.05$
and $\epsilon = 0.01$ for one $rns$-sample, together with 
$Gw^{(m)}-1$, where we substitute 1 for $K$ . 
For $\epsilon=0.05$ it is difficult, due to fluctuations in $w^{(m)}$,
to see that $\left[g^{(m)}\right]^2$ and $Gw^{(m)}-1$ are comparable. 
When we take the less value $\epsilon = 0.01$,
however, we can ensure that $\left[g^{(m)}\right]^2 \sim Gw^{(m)}-1$.
In Fig.~10 we plot $w^{(m)}$. We see it is quite low ($<$ 0.2) 
when $\epsilon =0.05$, while it increases to $ \sim 0.6$ with 
$\epsilon=0.01$.
Figure~11 shows $\langle \! \langle C^{(m)} \rangle \! \rangle$ over 
100 $rns$-samples, which should be compared with $E_0$ plotted by a
dashed line.
We see that when $\epsilon = 0.01$
values of $\langle \! \langle C^{(m)} \rangle \! \rangle$ 
after the transient decrease are nearly equal to $E_0$.
In addition, it is noticeable that data with $\epsilon = 0.05$, whose
overlaps with the exact eigenstate ($w^{(m)}$) are less than 0.2, are also
closely located to $E_0$ when $m > 10$.

\section{Numerical Results on Large Systems}

In this section we study the $J_1-J_2$ model on large lattices,  
the 36-site $J_2=-1$ model with $l=18$ in (\ref{j1j2q4}) and the 64-site
$J_2=0$ model with $l=0$.
Here we measure only $C^{(m)}$ and 
\begin{eqnarray}
\langle \psi^{(m-1)} \mid \left[M_{\{\eta^{(m)}\}}\right]^2 \mid 
\psi^{(m-1)} \rangle \sim 1+ \left[g^{(m-1)}\right]^2  \ ,
\label{neqggp1}
\end{eqnarray}
for $m=1,2,\cdots,m_{\rm max}$.  
The quantity (\ref{neqggp1}) is measured to make sure that 
$\left[g^{(m-1)}\right]^2$ does not diverge for large values of $m$. 

First we show results for the 36-site $J_2=-1$ model.
Since we do not know the exact eigenstate of
this system, we employ the N\'{e}el state as the initial trial state 
$\mid \psi^{(0)} \rangle$. For a few hundred of $rns$-samples we calculate 
$C^{(m)}$ and $\left[g^{(m-1)}\right]^2$ with $m_{\rm max}$=1000 with several
values of $\epsilon$ between $0.02$ and $0.1$.
For any of the $rns$-sample we observe $\left[g^{(m)}\right]^2
\sim 6.3 $ when $m \gtrsim m_0 = 200$ regardless of the value of $\epsilon$. 
The upper bound of the number of basis states whose coefficients are 
non-zero ranges between 
$\sim 8 \times 10^2$ (for $\epsilon = 0.1$) and 
$\sim 1.9 \times 10^4$ (for $\epsilon = 0.02$) in the expansion of
$M_{\{\eta ^{(m)} \}} \mid \psi^{(m-1)}\rangle$, while it is within 
$\sim 6 \times 10^4$ (for $\epsilon = 0.1$) and 
$\sim 1.4 \times 10^6$ (for $\epsilon = 0.02$)
in the expansion of
$\hat H M_{\{\eta ^{(m)} \}} \mid \psi^{(m-1)}\rangle$.  
Figure 12 and 13 plot $\langle \! \langle C^{(m)} 
\rangle \! \rangle _{n_{\rm smpl}}$ with $n_{\rm smpl}=100$ for several
values of $\epsilon$. Dashed lines in these Figures present 
the value of $E$ obtained 
from ref.~4, which amounts $E_0=58.659$ for this system. 
We see that the data for $m \gtrsim m_0$ are near $E_0$ 
with some fluctuations.
Let us denote the normalization factor for $\mid \psi ^{(m)} \rangle$ 
obtained in the $k$-th sample by $C_{\{\eta\}_k}^{(m)}$ as before.
In order to estimate the exact eigenvalue of $E$ based on (\ref{estime})
with a better precision, we take an average of $C_{\{\eta\}_k}^{(m)}$ 
not only over $k$ but also over $m=m_0$, $m_0+\Delta m$, $m_0+2\Delta m$,
$\cdots$, with some positive integer $\Delta m$. Namely, we calculate 
\begin{eqnarray}
\overline{C^\gamma} \equiv \frac{1}{(n_{\rm max}+1) \cdot n_{\rm smpl}}
\sum_{k=1}^{n_{\rm smpl}} \sum_{n=0}^{n_{\rm max}} \left[C_{\{\eta\}_k}^
{(m_0+n\Delta m)}\right]^{\gamma}  \ ,
\ \ \ \ \ n_{\rm max} \equiv \left[\frac{m_{\rm max}-m_0}{\Delta m}\right] \ ,
\label{cbarim}
\end{eqnarray}
for $\gamma=1,2$ and 
\begin{eqnarray}
Er \equiv 2 \sqrt
{\frac{\overline{C^2}-\overline{C}^2}
{{(n_{\rm max}+1)\cdot n_{\rm smpl}}-1}} \ . 
\label{errim}
\end{eqnarray}
It should be noted, however, that if $\Delta m$ is not large enough  
there will be a correlation between 
$C_{\{\eta\}_k}^{(m)}$ and $C_{\{\eta\}_{k}}^{(m+\Delta m)}$, 
in contrast to
that $C_{\{\eta\}_k}^{(m)}$ and $C_{\{\eta\}_{k'}}^{(m')}$ 
are statistically independent by definition if $k \neq k'$.
We therefore determine $\Delta m$ by the 
two-sided $5\%$ $t$-test of the hypothesis that there is no correlation 
between $C_{\{\eta\}_k}^{(m)}$ ($m \geq m_0$) and  
$C_{\{\eta\}_k}^{(m+\Delta m)}$. The result of the test suggests that  
$\Delta m$ should be more than 90, so we set  $\Delta m = 100$ for this 
model.
The results thus obtained with $n_{\rm smpl} = 100$ $(200)$ for 
$\epsilon < 0.07 $ $(\geq 0.07) $ are shown in Fig.~14. 
We observe that they are located close to $E_0$ but small 
discrepancies, which seem to be linearly dependent of $\epsilon$, 
still remain. 
So we carry out the weighted fit by the method of lease squares,
assuming that 
\begin{eqnarray}
\overline{C} = C_0 - A\epsilon \ ,
\label{lsfite}
\end{eqnarray}
where $C_0$ and $A$ are positive constants to be pursued.
Results of the fit using data with $0.02 \leq \epsilon \leq 0.1$ 
($0.02 \leq \epsilon \leq 0.06$) are 
$C_0 =  58.644 \pm 0.013 $ ($C_0 = 58.673 \pm 0.014 $). 
We plot the fitted line as well as 
the error of mean square for $C_0$, which we denote by $\Delta C$,
by solid (dashed) lines in the Figure. 
These values lead, since $l=18$ and $E_0=58.659$ for this system,  
$|\Delta C_0/(C_0-l)| = 0.00032 $ $(0.00034)$, or 
$|\{(E_0-l)- (C_0-l)\}/(E_0-l)| = 0.00037$ $(0.00034)$,
in the fit using data between $\epsilon = 0.02$ and 0.1 (0.06).
It is thus evident in this model that $C_0$ gives a good estimate for
$E$, which strongly support the realization of the RSSS equilibrium 
in this large-sized system.   

Similar investigations 
are made on the $l=0$, 64-site $J_2=0$ model, 
for which the most accurate value of $E$ is known to be 
$E_0 = 43.107$\cite{square}.
From several tens of $rns$-samples with $m_{\rm max} = 300$ or 
$m_{\rm max} = 500$ starting from the approximate state used  
in ref.~1 we obtain the following results. 
The RSSS equilibrium seems to be realized for $m \gtrsim m_0 = 200$,
where  $\left[g^{(m)}\right]^2 \sim 32$ for all values of  
$\epsilon$ we measured. 
The number of basis states whose coefficients are non-zero is, 
for $m \gtrsim m_0$,  
$\sim 1.3 \times 10^4$ ($\sim 3.5 \times 10^4$, $\sim 7.9 \times 10^4$,
$\sim 3.1\times 10^5$) in the expansion of
$M_{\{\eta ^{(m)} \}} \mid \psi^{(m-1)}\rangle$ for $\epsilon = 0.05$
(0.03, 0.02, 0.01), while it is 
$\sim 9.5\times 10^5$ ($\sim 2.6 \times 10^6$, $\sim 5.9 \times 10^6$,
$\sim 2.3 \times 10^7$) in the expansion of
$\hat H M_{\{\eta ^{(m)} \}} \mid \psi^{(m-1)}\rangle$.
From the two-sided $5\%$ $t$-test
we conclude that $\Delta m = 20$ is enough for 
$C_{\{\eta\}_{k}}^{(m+\Delta m)}$ to have no correlation with 
$C_{\{\eta\}_k}^{(m)}$ $(m \geq m_0)$.
Figure 15 plots values of $\overline{C}$ obtained from $(n_{\rm max}+1)
\cdot n_{\rm smpl}$ data, namely $6 \times 100$, 
$16 \times 10$, $6 \times 50$ and $6 \times 36 $ data 
with $\epsilon = 0.05$, $0.03$, $0.02$ and $0.01$, respectively. 
Results from the least square fit to 
(\ref{lsfite}) are also shown in the Figure. 
The fitted value of $C_0$ is $43.099 \pm 0.025$, 
which agrees well with $E_0$. 
With these values we obtain $|\Delta C_0/C_0| = 0.00058$, or 
$|(E_0- C_0)/E_0| = 0.00019$. 
This provides another example to indicate that the RSSS equilibrium is
established and that our way to estimate $E$ in the equilibrium is 
quite powerful. 

\section{Summary and Discussions}

In this paper we numerically study quantum spin systems with 
positive definite Hamiltonians by means of the RSSS method. 
We find that a kind of equilibrium, which we call the RSSS
equilibrium, exists.
We also notify that in this equilibrium we can effectively estimate 
the energy eigenvalue of the ground state.

In the RSSS method we
recursively calculate the $m$-th normalized intermediate state 
from the $(m-1)$-th one, starting from an initial trial state. 
The procedure is as follows.
First we operate the $m$-th random choice matrix to the $(m-1)$-th
state, which drastically reduces the effective size of the vector space. 
Then we successively operate the Hamiltonian, which again increases 
the number of the basis states relevant to the state. 
Finally we normalize the resultant state to obtain the $m$-th
normalization factor and the $m$-th normalized intermediate state.
What we assert is that, 
after repeating this procedure many times, the $m$-th
state comes to contain a finite portion of the ground state which is 
irrelevant to $m$. This means that the system is in the RSSS
equilibrium.

Our results on the 16-site 
and 36-site $J_1-J_2$ model with couplings $J_1=1$ and $J_2=-1$ 
as well as the 24-site and 64-site model with $J_1=1$ and $J_2=0$
afford abundant evidence of the RSSS equilibrium in these systems.
In addition, from the normalization factors stated above 
we obtain satisfying results which estimate the ground state 
energy on large lattices within $0.04\%$ of precision.
Our study on small and large lattices with various values of 
the parameter $\epsilon$ suggests that the RSSS equilibrium is observable 
for any value of $\epsilon$, as far as it keeps the 
random choice matrices non-zero.

A few remarks are in order.
In the present study we are solely concerned with systems having 
positive definite Hamiltonians, to which many useful methods are known 
to calculate their energy eigenvalues. In fact, a large amount of numerical
work for these systems has been already reported\cite{book1,book2}. 
What is the merit of our way to estimate the energy eigenvalue, then?
A remarkable feature is that it is very simple, having little to do with 
details of the system such as the dimensionality and so on.
Even for more complicated Hamiltonians our way to study the 
RSSS equilibrium is as simple as for those studied here.

How shall we think about the systems whose Hamiltonians are not positive 
definite? We suppose they also realize the RSSS equilibrium when the 
parameter $\epsilon$ is small. Preliminary results on the frustrated 
$J_1-J_2$ model and the triangular Heisenberg model are promising. 

There remain some fascinating tasks.
The most important one is to give a rigorous proof of the RSSS
equilibrium from mathematical point of view.
Another task is to analytically evaluate fluctuations we neglected 
in this paper. For this purpose it is necessary to investigate
$\epsilon$-dependency of each dropped term in (\ref{wmp1ex}) and
(\ref{cmp1sq}). We expect that contributions from these terms 
result in the linear relation (\ref{lsfite}) for sufficiently 
small values of $\epsilon$.
Arguments how to extract physical quantities other than the energy 
eigenvalues in the equilibrium would be also necessary.

\begin{figure}[p]
\begin{center}
\scalebox{0.5}{\includegraphics{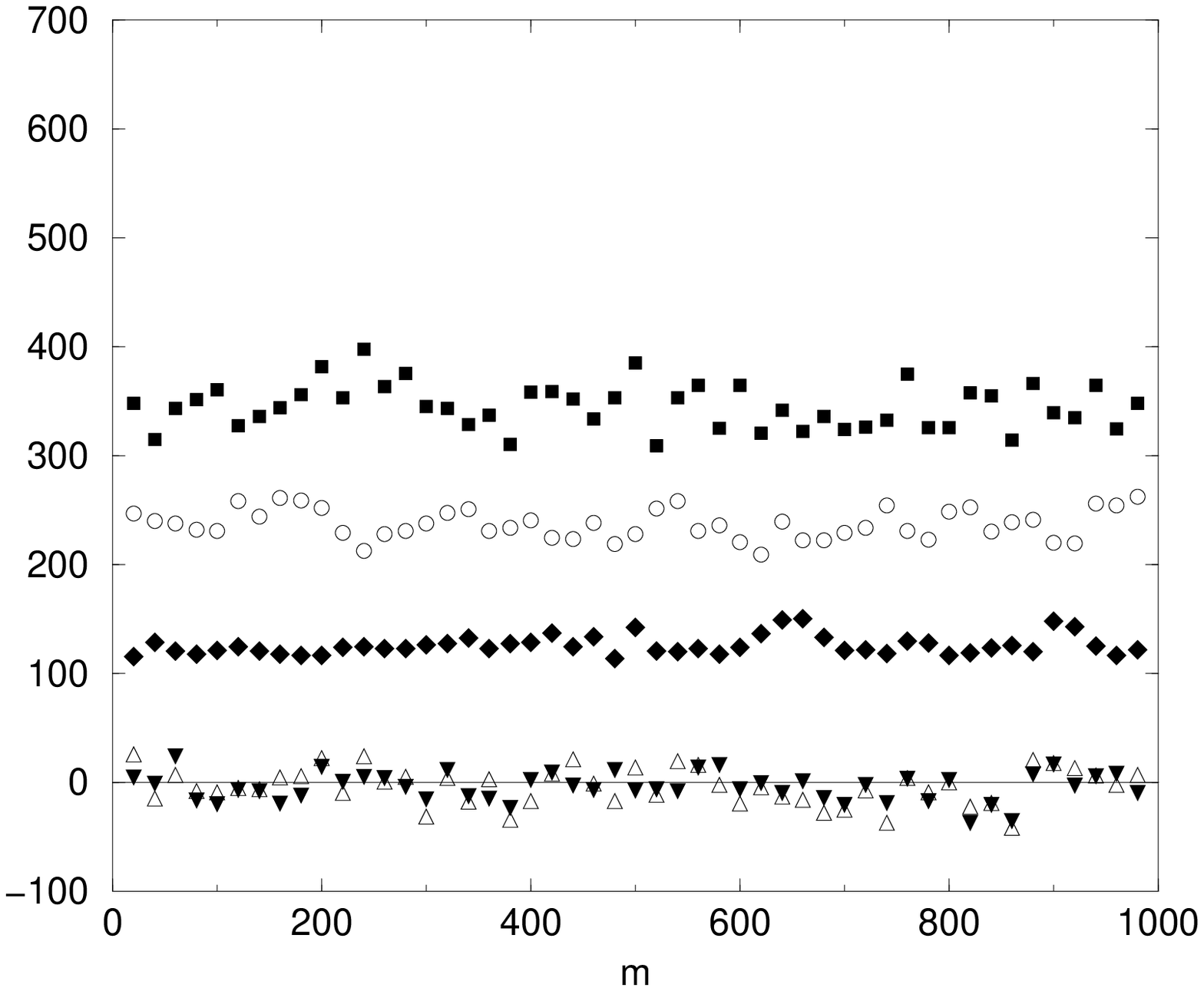}}
\caption{Terms in (\ref{cmp1sq}) for the $J_2=-1$ model with $\epsilon = 0.1$. 
Open circles are $E^2\left[w^{(m)}\right]^2$, 
filled diamonds are $\langle \chi^{(m)}\mid \hat H ^2 \mid \chi^{(m)} \rangle 
\left[g^{(m)}\right]^2$, 
filled squares are $\langle \zeta^{(m)}\mid \hat H ^2 \mid \zeta^{(m)} \rangle 
\left[s^{(m)}\right]^2$, open triangles are 
$2E^2w^{(m)}\langle \psi_{\rm E} \mid \chi^{(m)}) \rangle g^{(m)} $
and filled triangles are
$2s^{(m)}\langle \zeta^{(m)}\mid \hat H^2 \mid \chi^{(m)}\rangle g^{(m)}$. 
}
\end{center}
\end{figure}

\begin{figure}[p]
\begin{center}
\scalebox{0.5}{\includegraphics{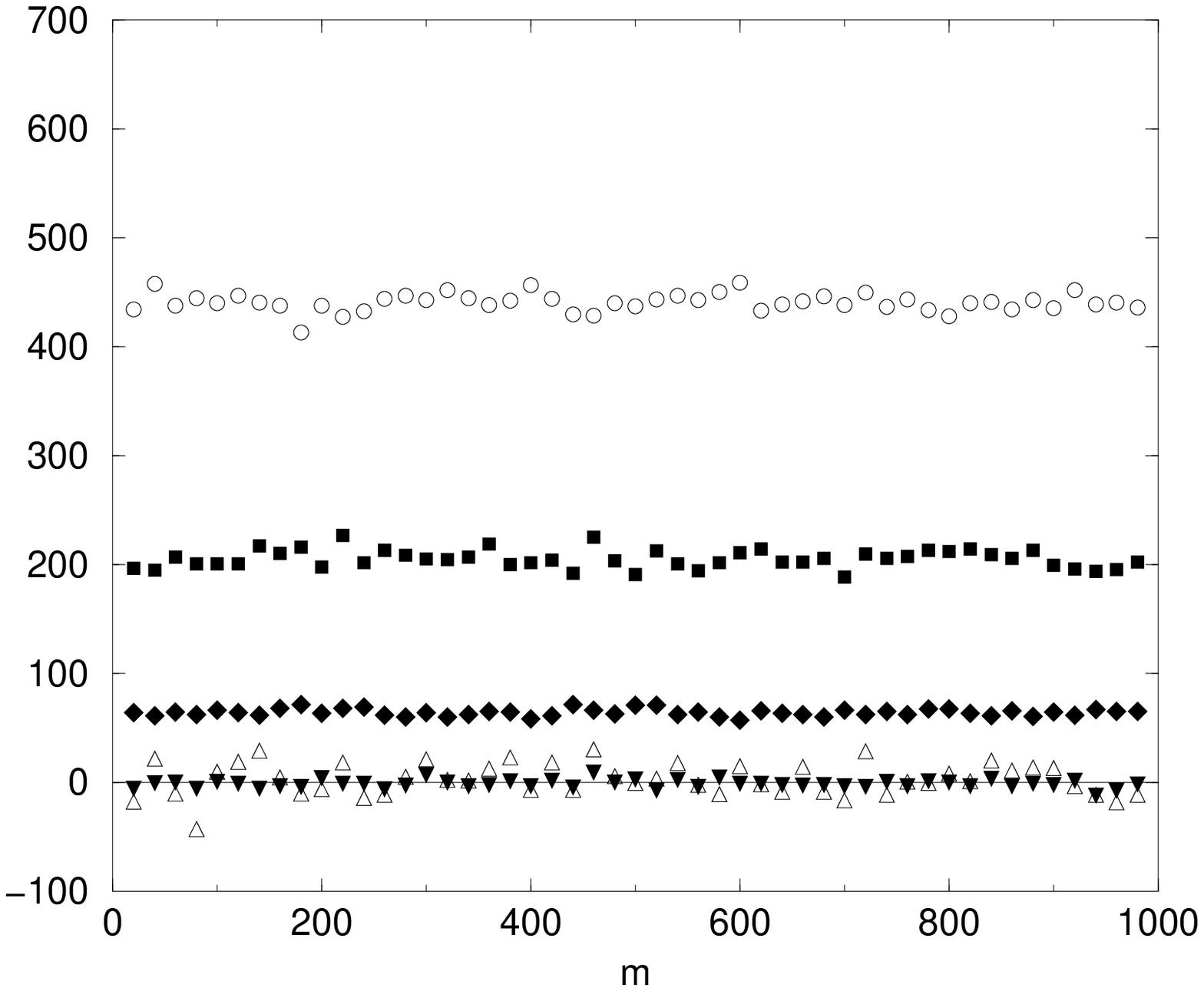}}
\caption{Terms in (\ref{cmp1sq}) for the 16-site $J_2=-1$ model with 
$\epsilon = 0.05$, where the same symbols as those in Fig.~1 are used.}
\end{center}
\end{figure}

\begin{figure}[p]
\begin{center}
\scalebox{0.5}{\includegraphics{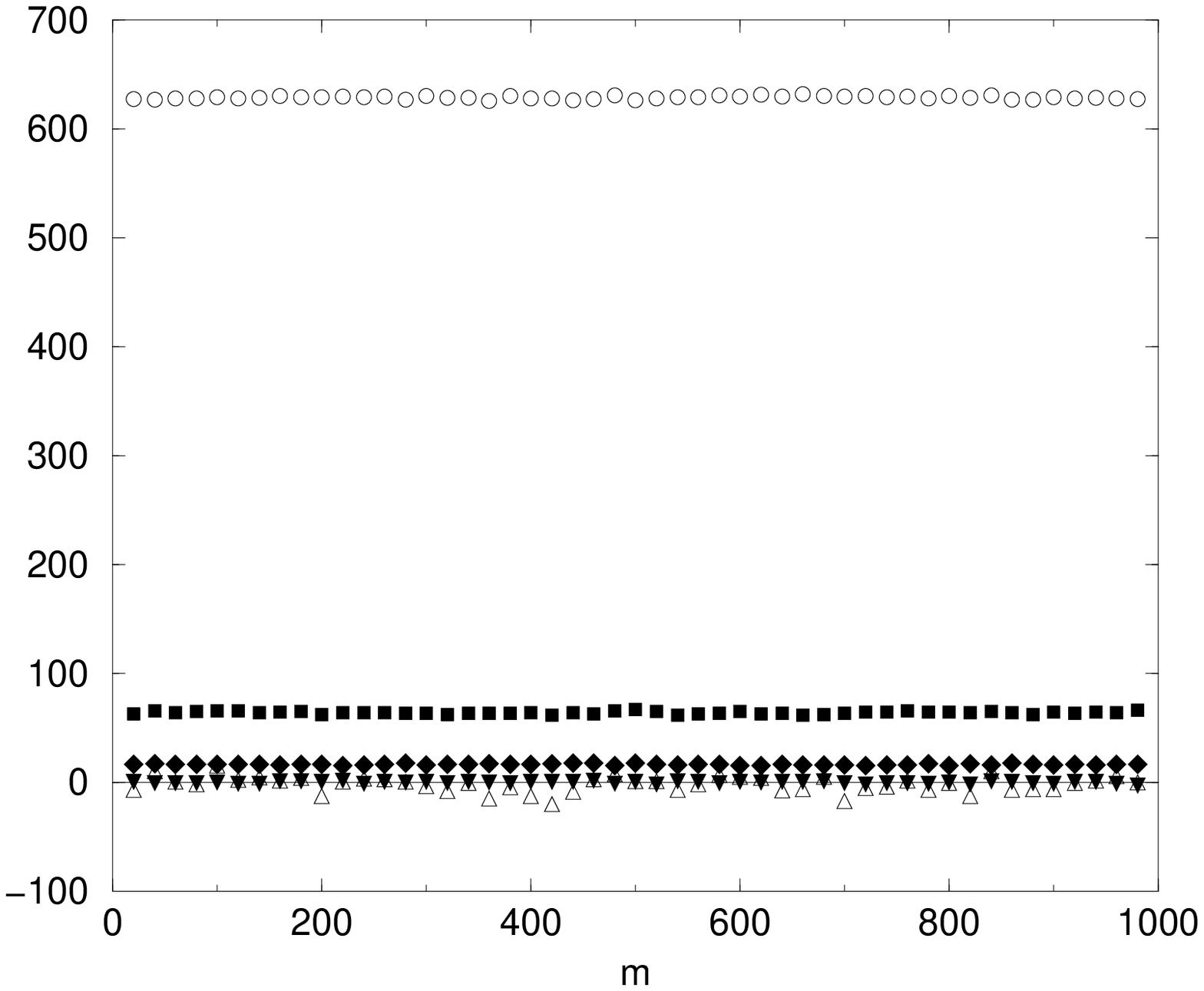}}
\caption{Terms in (\ref{cmp1sq}) for the 16-site $J_2=-1$ model with 
$\epsilon = 0.02$, where the same symbols as those in Fig.~1 are used.}
\end{center}
\end{figure}

\begin{figure}[p]
\begin{center}
\scalebox{0.5}{\includegraphics{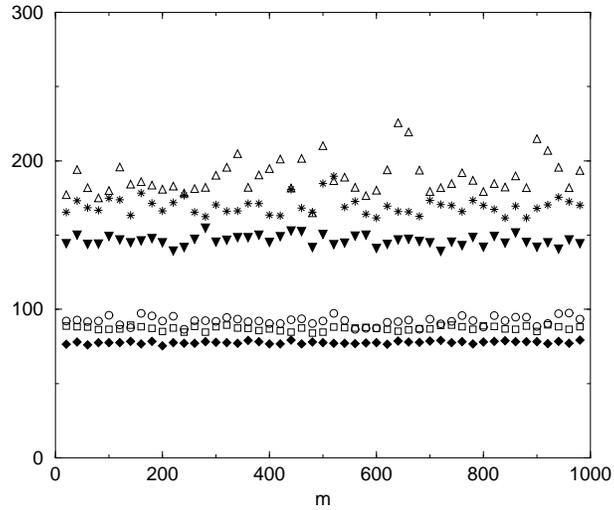}}
\caption{$\langle \zeta^{(m)}\mid \hat H ^2 \mid \zeta^{(m)} \rangle $
for the 16-site $J_2=-1$ model with 
$\epsilon = 0.1$ (open triangles), $\epsilon = 0.05$ (asterisks),
$\epsilon = 0.02$ (filled triangles) and 
$\langle \chi^{(m)}\mid \hat H ^2 \mid \chi^{(m)} \rangle $
with $\epsilon = 0.1$ (open circles), $\epsilon = 0.05$ (open squares),
$\epsilon = 0.02$ (filled diamonds).
Note that $\langle \zeta^{(m)}\mid \hat H ^2 \mid \zeta^{(m)} \rangle >
\langle \chi^{(m)}\mid \hat H ^2 \mid \chi^{(m)} \rangle $ always holds.}
\end{center}
\end{figure}

\begin{figure}[p]
\begin{center}
\scalebox{0.5}{\includegraphics{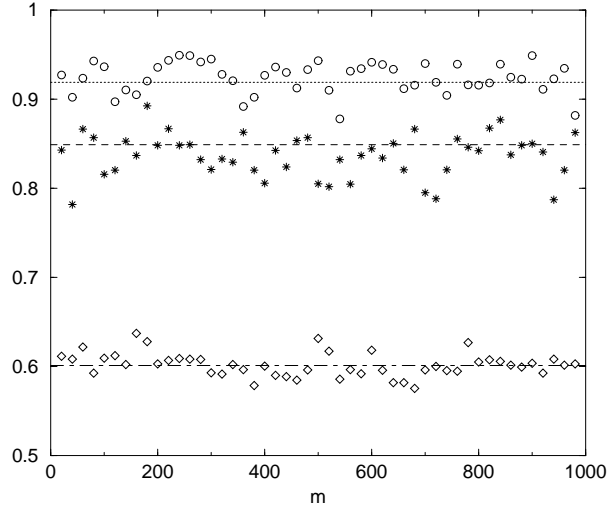}}
\caption{$\sum_{|c_i^{(m)}| \leq \epsilon} \left[c_i^{(m)}\right]^2$ 
for the 16-site $J_2=-1$ model with 
$\epsilon = 0.1$ (open circles), $\epsilon = 0.05$ (asterisks) and
$\epsilon = 0.02$ (open diamonds). We also show the value of $K$ 
estimated from last 200 data with $\epsilon = 0.1$ ($0.05$, $0.02$),
which is $K=0.919$ ($0.849$, $0.601$),  
by a dotted (dashed, dot-dashed) line, respectively.
 }
\end{center}
\end{figure}

\begin{figure}[p]
\begin{center}
\scalebox{0.5}{\includegraphics{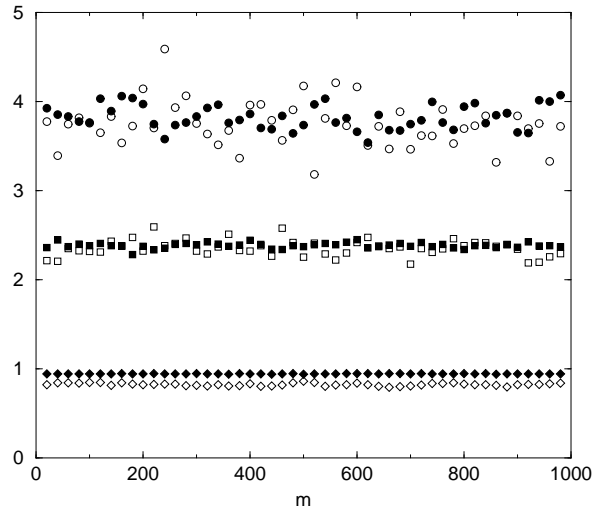}}
\caption{$[g^{(m)}]^2$ for the 16-site $J_2=-1$ model with
$\epsilon = 0.1$ (open circles), 
$\epsilon = 0.05$ (open squares), $\epsilon = 0.02$ (open diamonds).
We also plot  
$Gw^{(m)}-K$ with $\epsilon = 0.1$ (filled circles), 
$\epsilon = 0.05$ (filled squares) and $\epsilon = 0.02$ (filled diamonds),
where $G/\epsilon= \sum f_i = 82.08$ and we use values of $K$ 
estimated in Fig.~5.}
\end{center}
\end{figure}

\begin{figure}[p]
\begin{center}
\scalebox{0.5}{\includegraphics{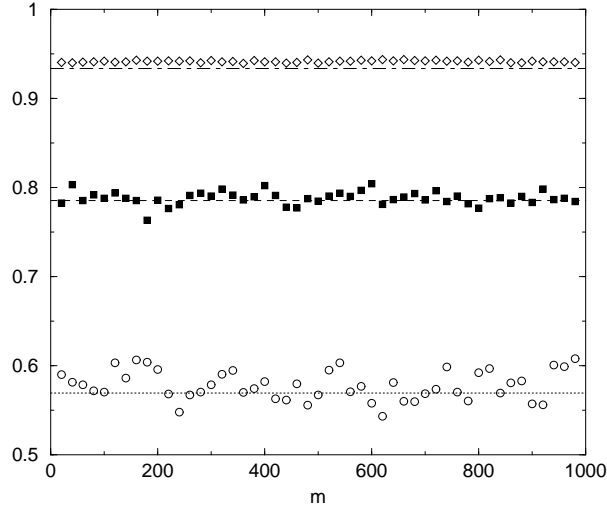}}
\caption{$w^{(m)}$ defined by (\ref{defwm}) 
for the 16-site $J_2=-1$ model with $ \epsilon = 0.1$ (open circles), 
$\epsilon = 0.05$ (filled squares) and $\epsilon = 0.02$ (open
 diamonds). Estimated solutions $w^{(eq)}$ for 
(\ref{solcw}) (a dotted line for $\epsilon = 0.1$, a dashed line for 
$\epsilon = 0.05$ and a dot-dashed line for $\epsilon = 0.02$) are also 
shown for comparison.}
\end{center}
\end{figure}

\begin{figure}[p]
\begin{center}
\scalebox{0.5}{\includegraphics{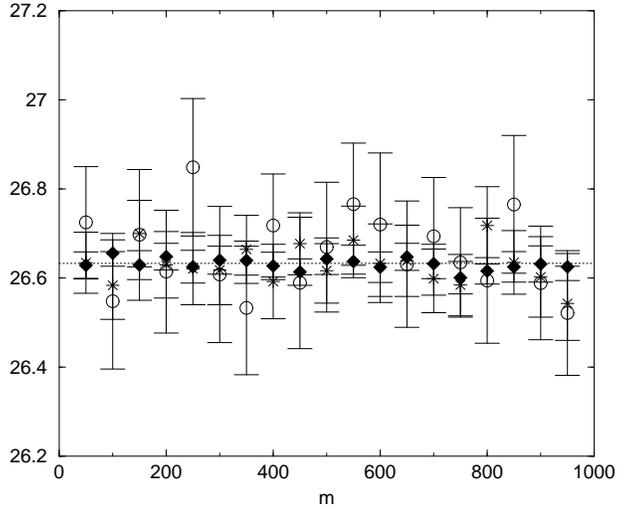}}
\caption{$\langle \! \langle C^{(m)} \rangle \! \rangle _{n_{\rm smpl}}$
($n_{\rm smpl}=100$) for the 16-site $J_2=-1$ model with
$\epsilon = 0.1$ (open circles), $\epsilon = 0.05$ (asterisks) and
$\epsilon = 0.02$ (filled diamonds). 
Errors shown in the figure are $Er^{(m)}$ defined by (\ref{errsmpl}). 
The dotted line indicates $E_0$, the exact value of $E$. }
\end{center}
\end{figure}

\begin{figure}[p]
\begin{center}
\scalebox{0.5}{\includegraphics{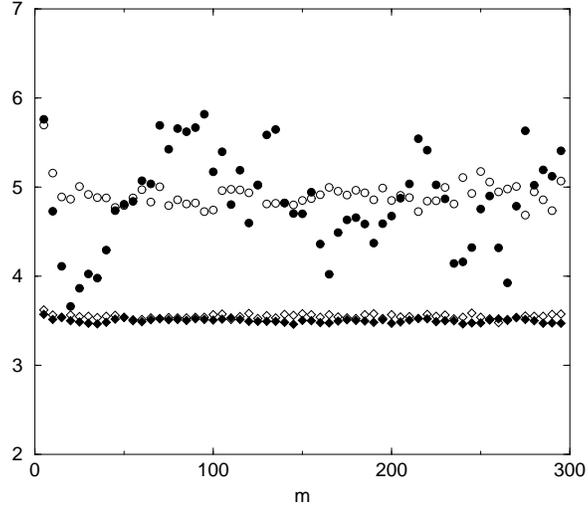}}
\caption{$[g^{(m)}]^2$ for the 24-site $J_2=0$ model with $\epsilon = 0.05$ 
(open circles), $\epsilon = 0.01$ (open diamonds). We also plot 
$Gw^{(m)}-1$ with $\epsilon = 0.05$ (filled circles), 
$\epsilon = 0.01$ (filled diamonds), where
$G/\epsilon= 792.1 $, for comparison.}
\end{center}
\end{figure}

\begin{figure}[p]
\begin{center}
\scalebox{0.5}{\includegraphics{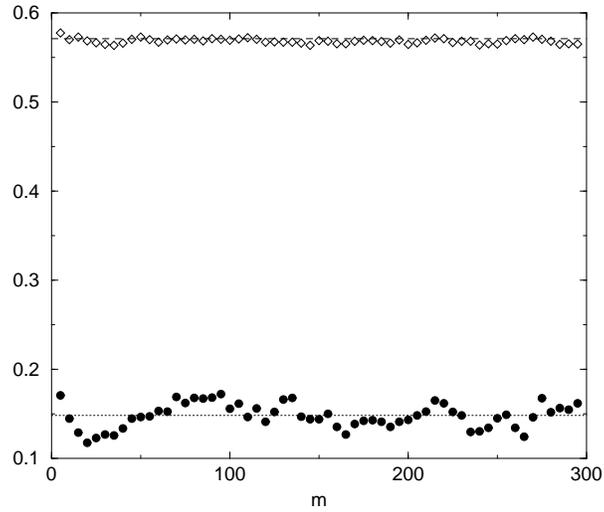}}
\caption{$w^{(m)}$ defined by (\ref{defwm}) 
for the the 24-site $J_2=0$ model with $\epsilon = 0.05$ (filled circles), 
$\epsilon = 0.01$ (open diamonds), together with estimated solutions 
$w^{(eq)}$ for (\ref{solcw}) represented by 
a dotted line ($\epsilon=0.05$) and a dashed line ($\epsilon=0.01$).
Here we put $K=1$ for both values of $\epsilon$.}
\end{center}
\end{figure}

\begin{figure}[p]
\begin{center}
\scalebox{0.5}{\includegraphics{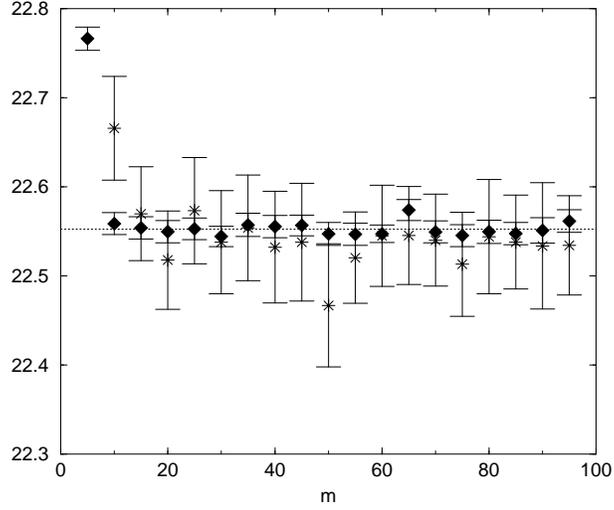}}
\caption{$\langle \! \langle C^{(m)} \rangle \! \rangle _{n_{\rm smpl}}$
($n_{\rm smpl}=100$) for the the 24-site $J_2=0$ model with $\epsilon = 0.05$
 (asterisks) and $\epsilon = 0.01$ (filled diamonds). 
$\langle \! \langle C^{(5)} \rangle \! \rangle _{n_{\rm smpl}}=24.01$  
when $\epsilon = 0.05$.  
Errors shown in the figure are $Er^{(m)}$ defined by (\ref{errsmpl}). 
The dotted line indicates the exact value of $E$.}
\end{center}
\end{figure}

\begin{figure}[p]
\begin{center}
\scalebox{0.5}{\includegraphics{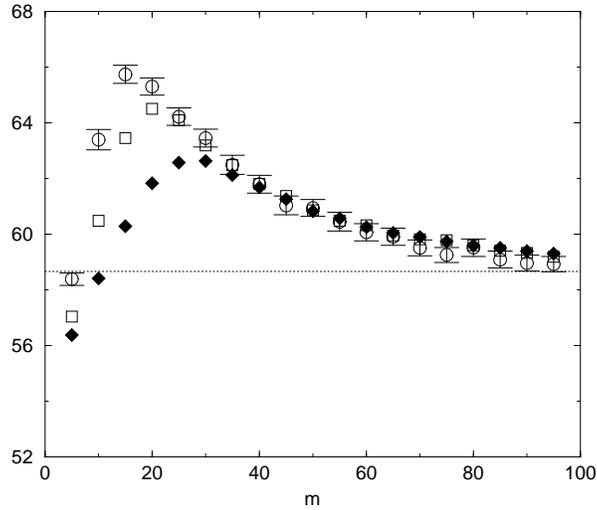}}
\caption{$\langle \! \langle C^{(m)} \rangle \! \rangle _{n_{\rm smpl}}$
($n_{\rm smpl}=200$)
for the 36-site $J_2=-1$ model for $m < 100$ with
$\epsilon = 0.1$ (open circles), $\epsilon = 0.05$ (asterisks) and
$\epsilon = 0.02$ (filled diamonds).
The initial trial state is the the N\'{e}el state.
Errors $Er^{(m)}$ defined by (\ref{errsmpl}) are within symbols when 
$\epsilon \leq 0.05$. A dotted line indicates $E_0$, 
the value of $E$ obtained from ref.~4. }
\end{center}
\end{figure}

\begin{figure}[p]
\begin{center}
\scalebox{0.5}{\includegraphics{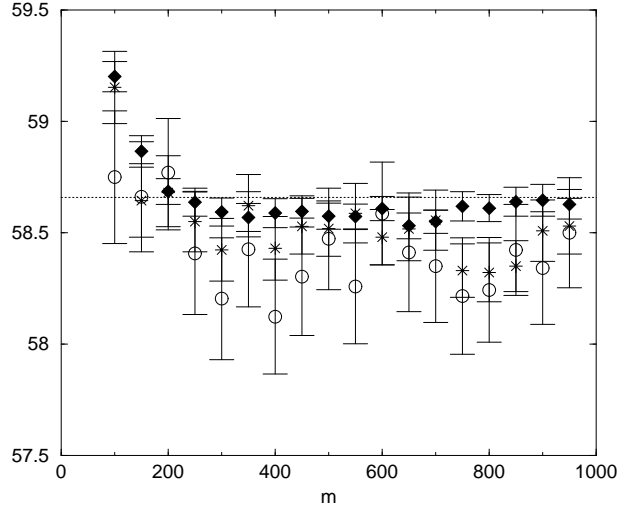}}
\caption{$\langle \! \langle C^{(m)} \rangle \! \rangle _{n_{\rm smpl}}$
($n_{\rm smpl}=100$) for the 36-site $J_2=-1$ model up to $m=1000$
calculated under the same conditions as Fig.~12, with  
the same symbols as those in Fig.~12. We show $E_0$ by a dotted line.}
\end{center}
\end{figure}

\begin{figure}[p]
\begin{center}
\scalebox{0.5}{\includegraphics{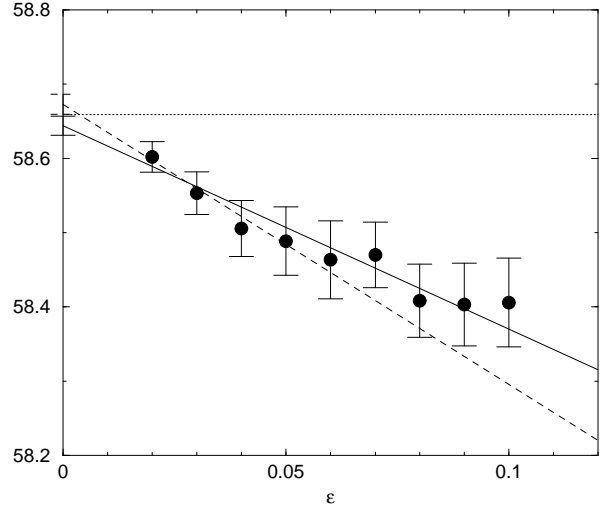}}
\caption{$\overline{C}$ versus $\epsilon$ calculated for the the 36-site
$J_2=-1$ model by (\ref{cbarim}) with $m_0=200$, $m_{\rm max}=1000$ and 
$\Delta m =100$. Errors are defined by (\ref{errim}).
The number of $C_{\{\eta\}_k}^{(m)}$'s used to calculate each 
$\overline{C}$ is $900$ $(1800)$ when $\epsilon < 0.07$ ($ \geq 0.07$).  
The dotted line indicates $E_0$. The solid (dashed) line
presents the result for the linear fit from the data with 
$0.02 \leq \epsilon \leq 0.1$ ($0.02 \leq \epsilon \leq 0.06$) 
obtained by the method of least squares assuming (\ref{lsfite}). 
Error bars at $\epsilon = 0$ show the error of mean square 
for the fitted $C_0$.}
\end{center}
\end{figure}

\begin{figure}[p]
\begin{center}
\scalebox{0.5}{\includegraphics{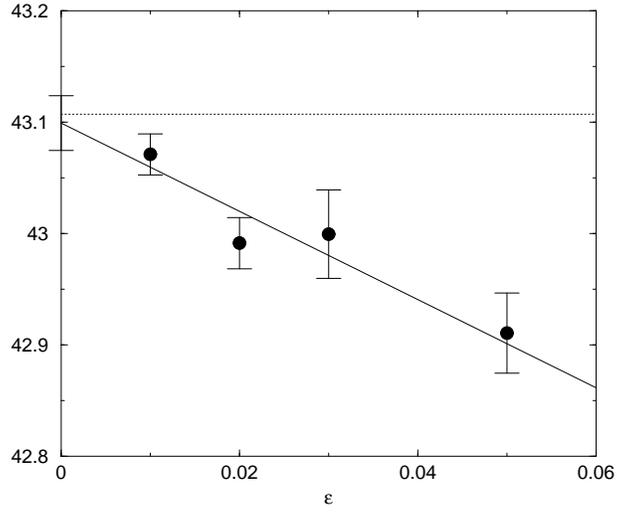}}
\caption{$\overline{C}$ for the the 64-site $J_2=0$ model versus 
$\epsilon$. We set $m_0 = 200$, $\Delta m =20$ in (\ref{cbarim}).  
Total number of data to calculate $\overline{C}$ is $600$ 
(160, 300, 216) when $\epsilon = 0.05$ (0.03,0.02, 0.01). 
Errors defined by (\ref{errim}) are also shown. 
The dotted line indicates the value of $E_0$. The solid line
presents the result for the least square fit by (\ref{lsfite}) 
with the error bar to show the error of mean square 
for the fitted $C_0$.}
\end{center}
\end{figure}


\begin{thebibliography}{99}
\bibitem{foot4}
Note that  $\mid \chi^{(m)} \rangle  g^{(m)}= 
\sum \mid i \rangle c_i^{(m)}(\eta_i^{(m+1)}-1)$ by definition, while 
neither $\mid \psi_ {\rm E} \rangle w^{(m)}$ nor 
$\hat H^2 \mid \zeta^{(m)} \rangle s^{(m)}$ includes any of 
$\{ \eta^{(m+1)}\}$ so that the statistical averages of
$w^{(m)} \langle \psi_{\rm E}\mid \chi^{(m)} \rangle g^{(m)}$ and
$s^{(m)} \langle \hat H^2 \zeta^{(m)}\mid \chi^{(m)} \rangle g^{(m)}$ 
over $\{\eta^{(m+1)}\}$ with a fixed 
$\{ \{\eta^{(m)}\}, \cdots,\{\eta^{(1)}\} \}$ are zero.
\bibitem{foot5} 
In numerical calculations, however, we should not choose large values of the 
parameter $\epsilon$ because they may generate a random choice matrix
whose diagonal elements are all zero.
\end{thebibliography}
\end{document}